\documentclass[preprint,12pt]{elsarticle}

\usepackage{amssymb}
\usepackage{amsmath}

\usepackage{comment}
\usepackage{color,soul}
\usepackage[table,xcdraw]{xcolor}
\usepackage{tabularx}
\usepackage{lscape}
\usepackage{url}
\usepackage{xcolor}

\journal{Applied Energy}
\begin{document}

\begin{frontmatter}



\title{Estimating the spatial economic and environmental impact of planned offshore wind energy in the USA using Environmentally Extended Multiregional Input-Output analysis 
}


\author[inst1]{Apoorva Bademi}
\author[inst2]{Miriam Stevens}
\author[inst3]{Isha Sura}
\author[inst1,inst2,inst3]{Shweta Singh}

\affiliation[inst1]{organization={Agricultural \& Biological Engineering},
            addressline={Purdue University}, 
            city={West Lafayette},
            postcode={47907}, 
            state={IN},
            country={USA}}

\affiliation[inst2]{organization={Environmental \& Ecological Engineering},
            addressline={Purdue University}, 
            city={West Lafayette},
            postcode={47907}, 
            state={IN},
            country={USA}}

\affiliation[inst3]{organization={Chemical Engineering},
            addressline={Purdue University}, 
            city={West Lafayette},
            postcode={47907}, 
            state={IN},
            country={USA}}

\begin{abstract}

{There is a substantial projected increase in energy generation from offshore wind farms in the US over the next three decades due to the rise in legislative commitments and funding from federal and state governments. 
Similar to other renewable energy technologies, the construction phase of offshore wind energy plants has potential environmental impacts and spill over effects to other regions, that need to be evaluated. Hence, developing offshore wind as a reliable domestic energy source requires a multiregional impact assessment of the economic and environmental spillover effects of constructing offshore wind farms in major lakefronts and coastal regions. Despite the lack of commercially operating offshore wind farms in the US, seven states have announced cumulative capacity commitments of over 28 GW by 2035. In this study, the spatial economic impacts of planned projects are estimated by combining NREL's Offshore Renewables Balance-of-system Installation Tool (ORBIT) for offshore wind energy with a multiregional input-output (MRIO) model of the United States built with the Virtual Industrial Ecology Laboratory (IE Lab). The ORBIT model provides the required capital investment for the installation of projects of interest which is then combined with MRIO model to evaluate spatial economic impact. 
Additonaly, the environmental impact occurring in different regions of the US due to spillover impacts of the construction of new wind farms was also evaluated. This was done using a newly developed multiregional GHG emissions dataset for the US IE Lab to estimate the supply chain emissions of constructing and installing offshore wind projects. The 5 projects considered in this research require a combined capital investment of \$16.3B and induces a total direct and indirect economic impact of \$27.6B from spillover effects within the US. The emissions results indicate that the states that involved in energy generation are highly impacted, emphasizing the fact that this can be improved by further decarbonizing the electricity grid. Based on the carbon pay-back period analysis, it takes these projects less than a year to offset all the emissions that were caused during the construction phase. The multiregional framework was used to identify which states experienced the most considerable spillover effects regarding emissions generation and economic activity required to support offshore wind projects.}

\end{abstract}


\begin{graphicalabstract}
\centering
\includegraphics[scale=0.6]{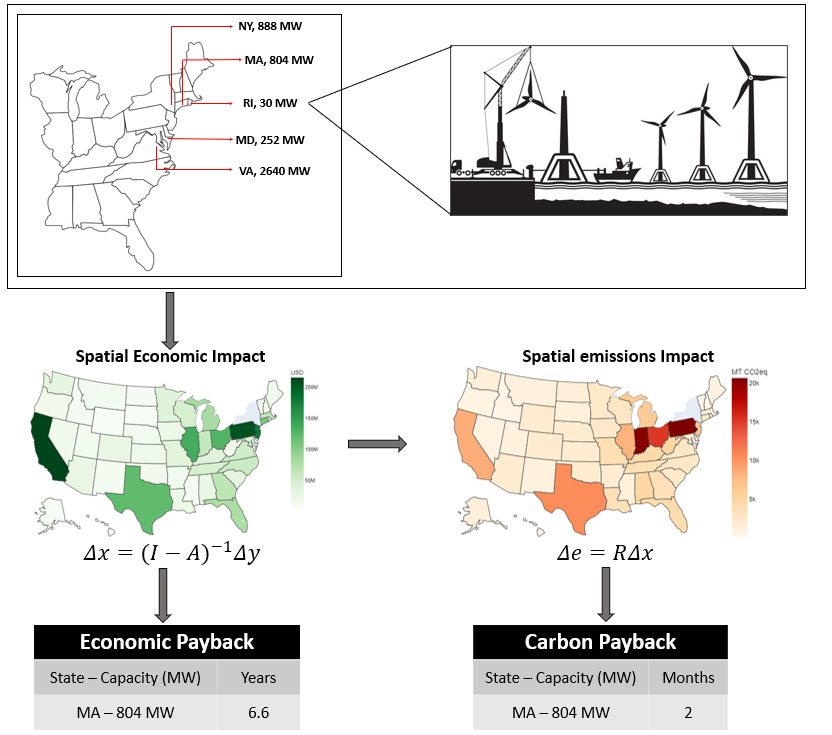}
\centering
\end{graphicalabstract}

\begin{highlights}
\item Spatial economic and emission impacts of offshore wind energy expansion in the US were determined using data from the US Industrial Ecology lab.
\item A new state-level GHG database was developed based on the US EPA inventory data which was used to estimate the spatial environmental impacts of offshore wind energy projects.
\item A total economic throughput of \$ 27 Billion is observed in the US due to installing 5 offshore wind plants.
\item The economic payback period of these projects ranges between 5-15 years. 
\item The carbon payback periods of the projects considered are less than a year, which is significantly smaller than their average lifetimes.
\end{highlights}


\end{frontmatter}


\section{Introduction}
\label{sec:sample1}
Decarbonizing energy sources is critical to achieving global CO\textsubscript{2} reduction targets to mitigate the impact of climate change. These targets, paired with rapid decreases in cost for renewable energy technologies over the past decade \cite{international_renewable_energy_agency_renewable_2022}\cite{international_renewable_energy_agency_renewable_2022-1}\cite{ro_renewable_nodate} have led to exponential growth in solar and wind energy worldwide; 970 TWh of solar generation \cite{IeaSolar} and 1528 TWh of wind generation \cite{IeaWind} have been added globally in the last decade. In addition to increasing affordability \cite{SEIA} \cite{Energy.govWindCost} \cite{LandBasedWindMarketReport}, the adoption of renewable energy sources has been further bolstered by their positive impact on the economy and society documented in several communities adopting renewable energy \cite{Shahbaz2020TheIndex}.
Renewable energy projects also provide consistent income to local and state governments. Additionally, it is estimated that these projects have created over 13,000 jobs in the U.S.' construction,  operation, and maintenance sectors between 1996 and 2019   \cite{SpringerKeyLands}. The share of renewable energy in the total energy supply was about 22\% for the U.S. in 2022, and was 37\% in Europe, 29\% in China, and around 28\% at the global level in 2021 \cite{Iea-Data}. 
Between 2010 and 2020, the USA added 82 G.W. of wind energy \cite{EnergyWindUSA}. Still, the share of offshore wind energy in the U.S. energy portfolio lags significantly behind Europe and other nations with leading renewable energy portfolios around the world \cite{musial_offshore_2021}. Given the extensive coastline of the U.S., offshore wind energy is promising as a reliable source of electricity generation compared to other renewable energy systems. It relies on wind over open waters, which have substantially greater speeds than onshore wind, allowing for high energy generation capacities with fewer turbines. Reaching speeds of 8.5 m/s to 10 m/s, wind speeds are greatest along the Northeast and West coast and the Great Lakes region \cite{musial_2016_2016}, which would enable convenient access to electricity for densely populated areas. With abundant space for offshore wind project installation in deep waters far from the shore along nearly all of the U.S. coastline \cite{wind_energy_technologies_office_new_2022}, the minimal use of land and interference with human activity makes this an even more promising system that could consistently supply energy to the grid.


In light of previous plans by the U.S. federal government to more aggressively integrate offshore wind energy into the national grid mix \cite{the_white_house_fact_2022}, there is a growing need for a proactive understanding of economic and emissions impacts of domestic offshore wind project deployment. Offshore wind technology is advancing rapidly with larger turbine capacities and a wider variety of floating and fixed-base anchor systems becoming commercially available \cite{jiang_installation_2021}. These advancements will enable building farms farther from shore with larger generating capacities, making offshore wind systems more efficient. Existing studies have considered the economic and energy impact of wind farm installations in the U.S., but those focused on offshore wind do not include broader economic impacts and Scope 3 emissions caused by project installation, and those considering impacts from an economy-wide perspective are focused on onshore systems. For example, the expected levelized cost of energy (LCOE) and annual energy production for a range of offshore turbine sizes and total plant capacities have been estimated by Shields et al., 2021 \cite{shields_impacts_2021}.
Similarly, a regionally specific analysis analyzing LCOE, infrastructure requirements, and trade-offs, and potential energy generation capacity was performed to evaluate the potential to implement offshore wind around Hawaii \cite{shields_cost_2021}. {Technoeconomic cost models that report project performance in terms of metrics like LCOE provide insight into the economic impact of the activities directly required to install and maintain energy generation projects. Limitations resulting from this method’s project specific focus is that such analyses do not include assessment of environmental emissions associated with projects, so complementary assessments are needed to identify the environmental factors to be considered when developing generation projects. Also, analysis using these methods can inform but not directly provide insight into supply chain impacts of energy generation projects. }In addition to studies focused on {projects' direct economic}  potential, environmental assessments also exist, such as a cradle-to-grave life cycle assessment (LCA) of onshore wind turbines deployed in Texas \cite{alsaleh_comprehensive_2019}, which found the most impactful lifecycle stages were equipment manufacturing and the least impactful stage for this particular project was disposal at end of life. A 2002 review of 72 studies found that the wide range of energy and carbon emissions intensities that have been reported for wind energy turbines resulted from analysis of procedural differences as well as the existing energy mix of the country of manufacture, end-of-life treatment, and material choice for large components \cite{lenzen_energy_2002}.  Extensive investigations of the {direct} cost and environmental impact of the offshore wind installation process are available, including US-specific studies, but expanding the scope of impact analysis of offshore wind energy in the U.S. to include more of the supply chain would provide a more complete understanding of how installing offshore wind will impact other parts of the economy and how the technology will contribute to the overall GHG emissions footprint of the nation.

 

Life cycle and supply chain impact of developing offshore wind can be informed using a multiregional impact assessment that can help evaluate the economic and environmental spillover effects of constructing wind farms in major lakefronts and coastal regions. Multiregional input-output (MRIO) analysis enables mapping project-specific activities and monetary costs to the induced economy-wide activities required to support them. Environmentally extended input-output (EEIO) analysis also allows for first-order accounting of supply chain impacts like scope 3 emissions. There is a lot of overhead required to develop offshore wind projects and activities involved in project planning, not just infrastructure creation, have the potential to contribute non-negligible GHG emissions. The EEIO method is a widely accepted approach for quantifying these indirect emissions. MRIO and EEIO analysis has been used to evaluate the impacts of offshore wind energy in other countries \cite{lundie_global_2019, velez-henao_hybrid_2021} and onshore wind energy in the US \cite{faturay_using_2020-1}, but a multiregional analysis of offshore wind in the US is missing.
The global environmental supply chain impacts for four projects spread across Germany, Turkey, and the UK, and their corresponding energy and carbon emissions payback periods, were evaluated using MRIO analysis \cite{lundie_global_2019}. The first LCA of a wind energy plant in Colombia was completed using a hybrid IO method and also found that their modeled system impacts were most sensitive to changes in the project capacity factor \cite{velez-henao_hybrid_2021}. In addition, there have been a considerable number of other environmental impact case studies of wind energy projects that incorporate an IO-based approach \cite{lenzen_energy_2002}. The economy-wide impacts of installing additional onshore wind in US states with the highest existing capacity were evaluated by \cite{faturay_using_2020-1} who also estimated the increase in energy consumption that would accompany new wind installations across sectors. {The current analysis shares similarities with the referenced study on onshore wind energy, as both utilize an MRIO framework to quantify the impacts of wind energy installations. Both studies highlight spatial economic and environmental spillover effects, focusing on impacts across states or regions, and aim to provide policymakers with valuable insights on how wind energy projects contribute to regional economies and national decarbonization efforts. However, there are notable differences between the two. The previous study centers on onshore wind installations in the top 10 wind energy-producing states, while the current analysis focuses on offshore wind projects planned along various U.S. coastal states. Additionally, the earlier study evaluates the short-term economic impacts of 500 MW capacity additions, whereas the current study examines larger offshore projects ranging from 30 MW to 2.6 GW. Methodologically, the first study employs NREL’s JEDI Wind Model, while the current analysis incorporates the NREL ORBIT model for installation cost estimation and utilizes state-level GHG databases for environmental impact assessments. While the earlier study emphasizes immediate economic impacts and provides actionable insights, it lacks environmental analysis and carbon payback considerations. In contrast, the current analysis offers a comprehensive assessment, integrating both economic and environmental payback periods and capturing broader spatial impacts, particularly for large-scale offshore projects.} Economic and emissions payback periods have also been used elsewhere as relevant indicators of wind energy project sustainability \cite{lundie_global_2019, lenzen_energy_2002}. {IO studies analyzing offshore wind energy emissions exist but to our knowledge none focus on the US. Emissions intensities often vary by region \cite{bruneau_income_2023} and as the US seeks to onshore manufacturing of renewable energy systems, estimating expected emissions from US-based production is important because the effort to increase domestic manufacturing capacity is taking place along with the need to manage US-induced GHG emissions. Furthermore, understanding the domestic monetary impacts of offshore wind systems at a regional level could help demonstrate how the benefits of economic activity induced by coastal projects would be distributed across the country.} Quantifying the regional impacts beyond mere project installation would provide information that can be used by policymakers and other planners to inform their decisions about investing in cleaner energy. 

{The main methods used for economic analysis of offshore wind energy are technoeconomic cost models and economic input-output analysis, of which only IO can provide estimates of the monetary costs of the supply chain and associated emissions. Given regional differences in sectoral emissions and the push to develop domestic capacity to manufacture offshore wind energy equipment, there is a need to further understand the broader impact of US offshore wind expansion. These impacts include the regional monetary implications of onshoring turbine manufacturing and the supply chain emissions that cannot be estimated using process LCA and are not included in most project development assessments. In response to this research gap, our} study addresses the following questions: 
(1) What is the total economic impact that could be expected to occur in different US regions due to upcoming offshore wind energy projects? (2) How much GHG emissions will be generated in different regions from the installation of the upcoming offshore wind farms? (3) How long will it take for these projects to offset the economic and carbon costs of their construction? 

To address these questions, the spatial economic impacts of five planned offshore wind projects are estimated by combining a discrete event simulation model of offshore wind energy installation \cite{nunemaker_orbit_2020} with a multiregional input-output (MRIO) model of the United States built with the Virtual Industrial Ecology Laboratory (IELab). We consider projects off the coast of Virginia (VA), Maryland (MD), New York (NY), Rhode Island (RI), and Massachusetts (MA). We also {develop and apply a new} multiregional GHG emissions dataset for the US IELab to estimate the supply chain emissions of constructing and installing offshore wind projects. These emissions estimates provide a more complete estimate of the overall spatial impact of expanding offshore wind energy than accounting solely for direct project costs and emissions. Hence, these emission and economic estimates can be used to calculate the time it will take for the proposed cleaner energy projects to offset the environmental impacts of their construction compared to those from conventional fossil fuel-based energy in terms of pay-back period for total emissions and economic impact. The multiregional framework is also used to identify which states may see the greatest spill-over effects in terms of emissions generation and economic activity. 

The novelty of this analysis lies in the developing a 52-region GHG emissions database and its application in the offshore wind industry to quantify the regional impacts of the upcoming projects. When integrated with existing datasets such as ORBIT and MRIO, this highlights the state-level impacts of these planned projects. This analysis shows the detailed potential spatial upstream impacts of offshore wind projects, which have not been addressed in the existing literature. This helps in more targeted decision-making in the federal-level decarbonization effort.

\section{Materials and Methods}
\begin{figure}[h!]
    \centering
    \includegraphics[scale=0.2]{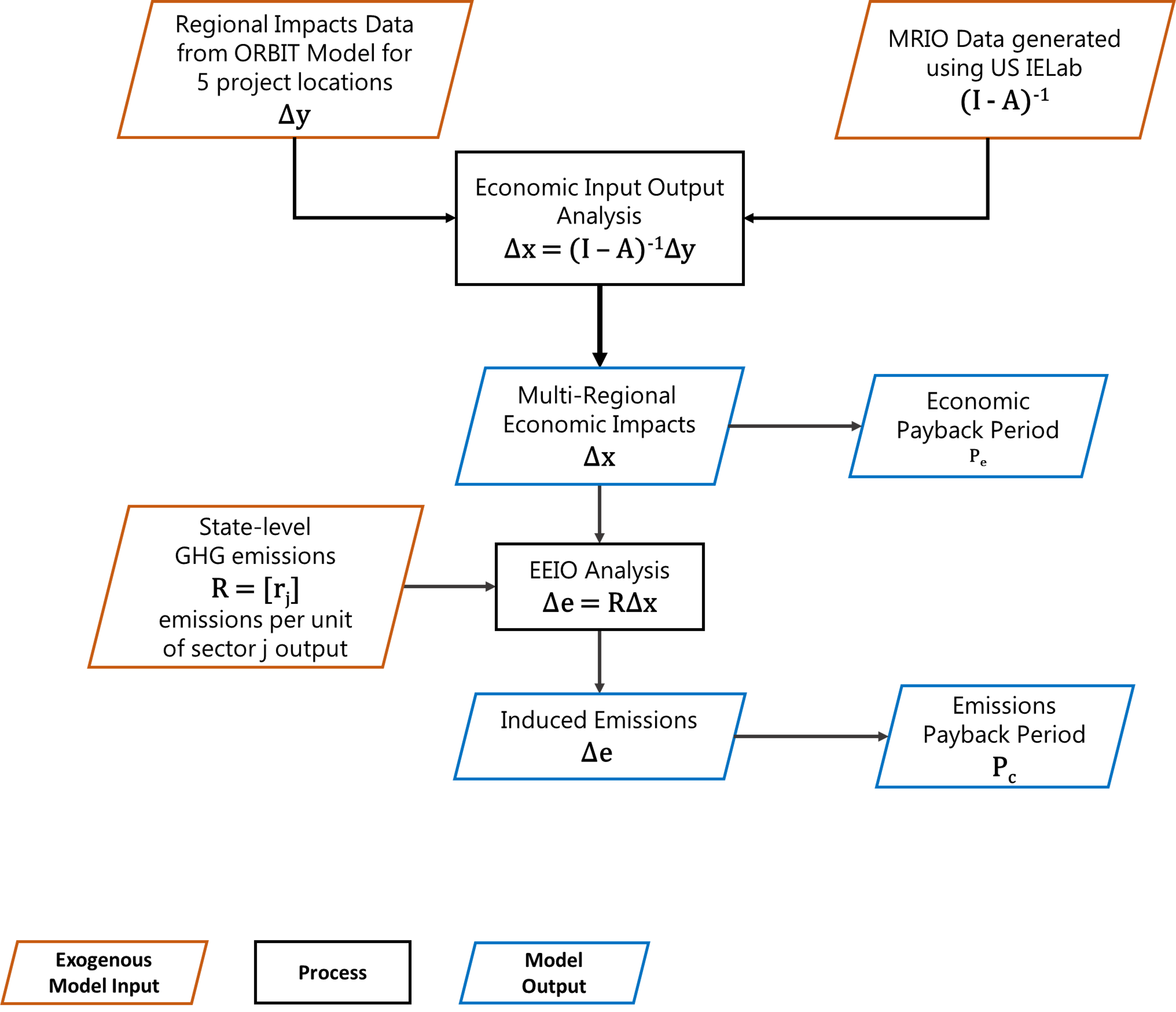}
    \caption{Multi-regional Economic and Emissions Impact Calculation Approach}
    \label{fig:ProjectWorkflow}
    \centering
\end{figure}

\begin{figure}[h!]
    \centering
    \includegraphics[scale=0.4]{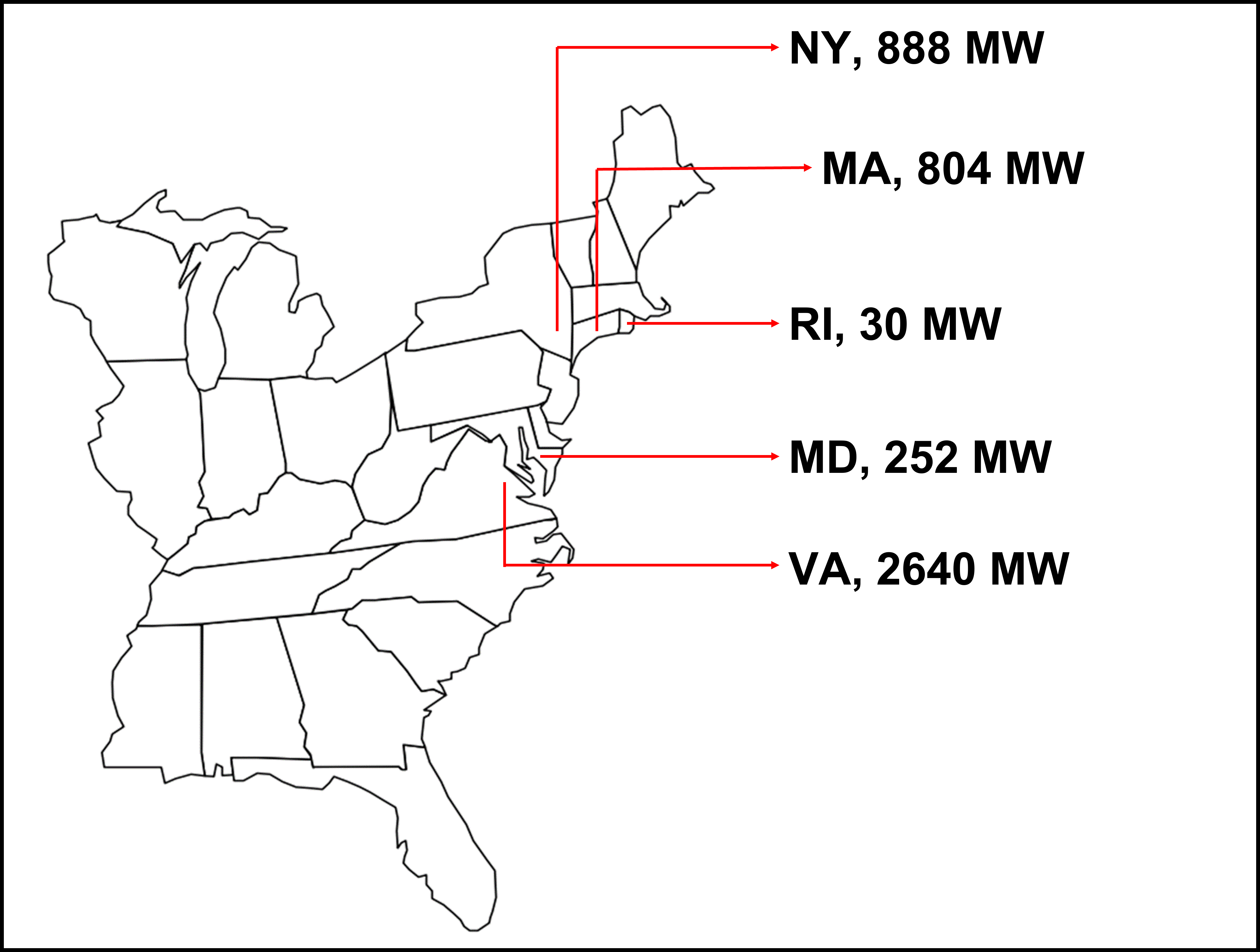}
    \caption{Planned Offshore wind projects considered for this study}
    \label{fig:Projects}
    \centering
\end{figure}

The method for this work entailed a) estimating the economic cost of construction and installation for expanding offshore wind energy capacity (Sec. \ref{localimpact}) and b) estimating the economic and emission impacts and payback periods using the US MRIO model (Sec. \ref{IOAnalysis}). The steps in quantifying the impacts are shown in Fig \ref{fig:ProjectWorkflow}. As the US MRIO model traces the spatial impact using the activities in economic sectors classified by the North American Industry Classification System (NAICS) code, a vital aspect of the methodology is connecting the local expense of each wind energy project to the MRIO model by mapping installation costs to the NAICS code for each sector in the region where the new wind energy plants are installed. 
The location and the capacities of the projects of interest analyzed in this work are shown in Fig \ref{fig:Projects}. The projects selected were representative of those currently planned to be constructed and operational by 2026. The capacity of the projects chosen for this analysis ranges between 30MW to 2.6GW. For reference, Hornsea 2 in the United Kingdom is presently the largest operational offshore wind with a capacity of 1.3GW \cite{Hornsea2}. {Offshore wind projects across various states in the U.S. are being developed and implemented by different agencies. Each project's capacity is obtained from the respective developer project information. The first offshore wind farm in the United States, located in Rhode Island, commenced commercial operations in 2016. This project was developed by the Coastal Resources Management Council (CRMC), a management agency who's responsibilities include the preservation, protection, development, and restoration of coastal areas \cite{Ri}. The MarWin 1 offshore wind project in Maryland is being developed by US Wind, Inc. This project is part of US Wind's broader Maryland Offshore Wind Project, which includes multiple phases to deliver clean, renewable energy to the state \cite{Offshore_MA}. Vineyard Wind 1, developed by Vineyard Wind LLC—a joint venture between Copenhagen Infrastructure Partners and Avangrid Renewables—is located in Massachusetts \cite{Barnstable}. The Empire Wind project, located offshore New York, is being developed by Equinor \cite{NYSERDA}. Dominion Energy is developing the Coastal Virginia Offshore Wind (CVOW) project. This project includes a pilot phase with two turbines aiming to generate 2.6 gigawatts of clean energy \cite{Dominion}.}

\subsection{Estimating Economic Cost of Offshore Wind Energy Expansion} \label{localimpact}
\label{sec:InstallationCost}
{It was first necessary to estimate the costs of installing offshore wind projects and map project cost categories to the economic sectors of the IO model.} The costs to build and install a given offshore wind energy project were estimated using the Offshore Renewables Balance-of-System and Installation Tool (ORBIT) developed by the National Renewable Energy Laboratory (NREL). The ORBIT model is an open-source, bottom-up design tool that computes balance-of-system (BOS) costs for hypothetical offshore wind energy projects using discrete event simulation \cite{nunemaker_orbit_2020}. In discrete event simulation, actors in a system are modeled as entities whose interactions can be discretized into a series of actions performed according to a predefined schedule\cite{borshchev_system_2004}. In the case of offshore wind installation as simulated by ORBIT, such an actor would be a wind turbine installation vessel that loads turbine substructures at the port, travels to the installation site, offloads and positions the substructure, moves to another turbine site, and repeats installation process, etc. BOS costs can be broadly categorized as falling under capital expenditure, installation, or development costs. Capital expenditure costs include those for the substructure connecting the wind turbine to the sea floor, array cables connecting the turbines, offshore substations that collect the generated electricity and transfer it to shore, and export cables that connect the offshore and onshore substations. Installation costs include those paid to installation vessels for their work and during weather delays and construction port fees. Development costs include site commissioning, decommissioning, and fees paid to the U.S. Department of Interior Bureau of Ocean Energy Management to obtain rights to develop the site, cover permitting, and complete surveys \cite{nunemaker_orbit_2020} .

Projects with highly customized design specifications can be simulated with the ORBIT model and the model scales costs with design parameters such as plant capacity, turbine rating, and site characteristics. This more accurate cost scaling with design parameters was the reason for using the ORBIT model instead of NREL’s Jobs and Economic Development Impacts (JEDI) model, which was used for a similar study analyzing onshore wind energy \cite{faturay_using_2020-1}. ORBIT contains default values representative of typical offshore wind projects. For fixed-base projects, this consists of wind turbines constructed in water at an average depth of 23 meters and within 35 km of landfall. These values ensure an accurate estimate of potential impacts from a general project. However, when data was available, these default values were replaced with site-specific factors such as project size and location characteristics (See table \ref{tab:design-specs}). Differentiation between the five projects modeled was mainly done through their planned total operational capacity, but data on sea depth at the installation site, distance from shore, and location mean wind speed were also included when available from project development reports. { Site-specific characteristics such as wind speeds over 7m/s, for all but the RI pilot project, and moderate depths are necessary features to make the selected sites developable, as described in reports for evaluating technical resource potential across the US \cite{wind_energy_technologies_office_new_2022} and for each project \cite{Ri}-
\cite{Dominion}, and are important for differentiating the installation and operating costs of each project. } The data in  Table \ref{tab:design-specs} was used to vary the input files from a default configuration for the ORBIT model. Project costs due to weather delays were estimated through the inclusion of site-specific weather data on hourly wind speed and wave height for a three-year time period from the ERA 5 dataset \cite{nicolas_era5_2023}; this weather data is used during the simulation to predict times when weather conditions are suitable for installation; when they are not, vessels pause work and delay costs are accrued. Project costs generated from the ORBIT model for each project were then mapped to the NAICS industry classification system used in the MRIO model. This mapping between cost category and NAICS codes, and the cost category descriptions are shown in  \ref{sec:ORBITCostMapping:appendix} and \ref{sec:ORBITCostCategoryDescription:appendix}, respectively. The project costs were used as a final demand vector \(\Delta \)y in the input-output analysis described next.

\begin{table}[]
\tiny
\caption{Design specifications used in ORBIT for the wind energy projects modeled. }
\label{tab:design-specs}
\begin{tabular}{lllllll}
\hline
 \textbf{\begin{tabular}[c]{@{}l@{}}Project \\ State\end{tabular}} &
  \textbf{\begin{tabular}[c]{@{}l@{}}Capacity \\ (MW)\end{tabular}} &
  \textbf{\begin{tabular}[c]{@{}l@{}}Turbine size \\ (MW)\end{tabular}} &
  \textbf{\begin{tabular}[c]{@{}l@{}}No. Turbines \\ Modeled\end{tabular}} &
  \textbf{\begin{tabular}[c]{@{}l@{}}Depth \\ (m)\end{tabular}} &
  \textbf{\begin{tabular}[c]{@{}l@{}}Distance to \\ landfall (km)\end{tabular}} &
  \textbf{\begin{tabular}[c]{@{}l@{}}Mean windspeed \\ (m/s)\end{tabular}} \\ \hline
RI & 36   & 12 & 3   & 24.4 & 4.8  & 6   \\
VA & 2640 & 12 & 220 & 22.5 & 43.5 & 8.5 \\
MA & 804  & 12 & 67  & 22.5 & 24   & 9.5 \\
NY & 888  & 12 & 74  & 32   & 22.2 & 7.5 \\
MD & 268  & 12 & 22  & 22.5 & 32   & 5.2 \\ \hline
\end{tabular}
\end{table}

\subsection{Direct and Indirect Impact Analysis using US MRIO} \label{IOAnalysis}
The total direct and indirect impact of the installation of these offshore wind projects was quantified using the US MRIO model. Here, the output of the ORBIT model is input to the US MRIO model as \(\Delta \)y for each region based on the region where installation is done.

\subsubsection{Economic Impacts of Expanding Offshore Wind Energy} 
\label{EconomicImpactCalc}
To calculate the economy wide impacts, we generated a US MRIO table at the 101 sector aggregation level using the US IE Lab \cite{faturay_using_2020-1}. This table includes 100 industrial sectors at a NAICS 3-digit sector level and a wind energy sector that is disaggregated from the utilities sector. US IE Lab is a  cloud-based platform for generating customizable input-output tables at various levels of sector and regional aggregation using an advanced optimization engine \cite{lenzen_compiling_2014}. We then derived an industry-by-industry direct requirements matrix ($A$) from the supply and use matrices using the Eurostat Model D method \cite{eurostat_eurostat_2008}, modified for multiregional SUTs as shown below. Transformation matrix $A$ was then converted to a compound total requirements matrix ($L$) where $L=(I-A)^{-1}$. Finally, the Leontief model in Eq. 1 was used to calculate the total economic impact ($\Delta x$), in units of million USD, using the modeled final demand ($\Delta y_r$) for each region\emph{r} obtained from the ORBIT model. In Eq. 1, $I$ is a unitless identity matrix with same dimensions as $A$.

\begin{equation} \label{eq1}
    \Delta x = (I-A)^{-1}\Delta y
\end{equation}

where A = 
$\begin{bmatrix}
A^{11} & A^{12} & . & . & . & A^{1r}\\
A^{21} & A^{22} & . & . & . & .\\
. & . & . & . & . & .\\
. & . & . & . & . & .\\
A^{r1} & . & . & . & . & A^{rr}
\end{bmatrix}$, $\Delta$x = 
$\begin{bmatrix}
\Delta x^1\\
\Delta x^2\\
.\\
.\\
\Delta x^r
\end{bmatrix}$, $\Delta y$ = 
$\begin{bmatrix}
\Delta y^1\\
\Delta y^2\\
.\\
.\\
\Delta y^r
\end{bmatrix}$


\bigskip
All project costs were assumed to be associated with the gross fixed capital formation category of final demand and with industries in the project state. We also assumed all project expenses, except turbine manufacturing, can currently be fulfilled with domestic manufacturing; these costs comprised the change to final demand vector $\Delta y_r$. While offshore wind turbines are currently imported, there is a national push to reshore their production. We also model the impacts from future turbine manufacturing as domestic expenses but calculate their impacts separately from other purchases to keep track of impacts from current versus future domestic manufacturing capacity. The future domestic turbine manufacturing is a key assumption in this analysis. The cost of the wind turbines was therefore considered a separate 'shock' to final demand and was differentiated from current domestic manufacturing capacity by including this expense in a separate final demand vector represented as $\Delta y_i$. 


\subsubsection{Emissions Impacts of Expanding Off Shore Wind Energy}

To calculate the emissions impact, we enhanced the US IE Lab with a satellite account that includes a newly developed multiregional GHG emissions dataset. 
This dataset is based on the inventory of US Greenhouse gas emissions and sinks published by the US Environmental Protection Agency (US EPA) \cite{US2021InventorySinks}. The inventory follows a comprehensive methodology for attributing the sources of emissions to different industries that is consistent with the United Nations Framework Convention on Climate Change (UNFCCC) format. The National Greenhouse Gas Industry Attribution Model uses this inventory to map the emissions to industrial sectors at the national level. However, the local impacts often vary widely from national averages due to vast differences in local economies. Hence, a new state-level GHG database was generated to estimate these impacts at a regional scale. The dataset comprises of GHG emissions by sector in 52 geographical regions of the US attributed to 100 industrial sectors as classified by the North American Industry Classification System (NAICS). {The major methodological novelty of this work is the development of this state-level EEIO satellite dataset.} Full details of the methodology followed for generating this dataset is detailed in \ref{sec:GHGMethod:appendix}.

Finally, a new wind energy sector was added to the existing 100-sector database to include the wind energy sector. This sector is assumed to have no emissions since the emissions occurring during the energy production through wind turbines are negligible. The new database for 101 sectors and 52 regions is then used in combination with the economic data to calculate the emissions factor for each sector. The emissions factor (EF) is defined  in Eq. 2 as the emissions generated per unit of industry output. 
\begin{equation} \label{eq2}
    ef = \frac{GHG emissions}{Total Industry Output}\
\end{equation}
where, 
ef is the emissions factor measured in (MT of CO\textsubscript{2}eq./Million Dollars)

The emissions factors are then multiplied element-wise with the economic impacts obtained as mentioned in Section \ref{EconomicImpactCalc} to enumerate the total emissions impact corresponding to each industrial sector's induced output in different regions (Eq. 3). 

\begin{equation} \label{eq1}
\Delta E = ef*\Delta x
\end{equation}

where $\Delta $E =
$\begin{bmatrix}
\Delta E^1\\
\Delta E^2\\
.\\
.\\
\Delta E^r
\end{bmatrix}$,
ef =
$\begin{bmatrix}
ef^1\\
ef^2\\
.\\
.\\
ef^r
\end{bmatrix}$ and $\Delta x$ =
$\begin{bmatrix}
\Delta x^1\\
\Delta x^2\\
.\\
.\\
\Delta x^r
\end{bmatrix}$

\subsection{Payback period analysis of new wind energy installation}
The expansion of offshore wind energy capacity is a huge and, for the US, novel infrastructure investment whose sustainability impacts might be usefully understood in a comparative context with the conventional energy it seeks to replace. Therefore we also calculate the economic and emissions payback periods for these projects. {By using MRIO analysis to include supply chain impacts in the total monetary and emissions cost of each project, which would not be known by project developers, the payback period includes a more complete accounting of life cycle costs.}

\subsubsection{Economic payback period}
The economic payback period, or break-even point, was calculated to measure the economic return on investment for the upcoming offshore wind projects. In this study, the payback period is determined using the following formula as described in a previous paper \cite{CharroufBetkaBecherifTabanjat+2018}:
\begin{equation} \label{eq2}
EPB = \frac{C_i}{(AEP*P_s)-C_{op}} 
\end{equation}
In Eq. 4, C\textsubscript{i} is the cost of initial investment, AEP is the net annual energy generation, P\textsubscript{s} is the average cost of electricity for the state in which the project is installed as reported by the Energy Information Administration \cite{EIAElectricityPrice}, and C\textsubscript{op} is the cost of yearly operation as obtained from the ORBIT model. The denominator represents the net revenue generation from the plant.

\subsubsection{Carbon payback period} \label{CBP}
The carbon payback period was assessed to determine the length of time the plant needs to be operational to offset the total emissions generated during the installation phase. The emissions avoided by displacing the energy from fossil fuels use are considered a kind of credit that may eventually outsize the emissions generated during the construction of the wind plants. In this scenario, the emissions occurring in the wind energy plant's use phase are assumed to be negligible compared to those occurring during the construction and installation phases. The lifetime of the wind energy plant is considered to be 25 years. The payback period is calculated using the method described below. Firstly, the avoided emissions rate (R\textsubscript{avoided}) is calculated in Eq. 5 as the difference in emissions intensities of each type of energy generation.   
\begin{equation} \label{eq3}
R_{avoided} = R_{avgelec} - R_{osw}
\end{equation}
where,

\noindent{R\textsubscript{avgelec}  is the emissions intensity of a conventional energy generation plant in MT of CO\textsubscript{2}eq/MW generated. The projected trend of decarbonization of the electricity grid for the next 25 years is considered when calculating the emissions intensity of conventional energy. We base our grid emissions forecast on the U.S. Energy Information Administration’s (EIA) Annual Energy Outlook 2020 projections, which provide grid decarbonization trends in the U.S. through 2050 \cite{USEIA2020}.}

\noindent{R\textsubscript{OSW} is the emissions intensity of the wind energy installation over its lifetime in MT of CO\textsubscript{2}eq/MW generated. It is calculated as the ratio of emissions generated during its lifetime to the total energy generated in its lifetime of 25 years. The lifecycle emissions in kg of CO\textsubscript{2}eq. per unit of energy by generation type was used to calculate the emissions for the 25 years.}

\noindent{The energy required to offset emissions (En\textsubscript{offset}) is calculated as the ratio of lifetime emissions generated by the wind energy plant to the avoided rate of emissions.}

\begin{equation} \label{eq4}
En_{Offset} = \frac{Em_{Lifetime}}{R_{avoided}}
\end{equation}

\noindent{Finally, the carbon payback period (CPB) is calculated as the time required to generate the required amount of energy to offset emissions.}

\begin{equation} \label{eq5}
CPB = \frac{En_{Offset}}{En_{Annual}}
\end{equation}

En\textsubscript{Annual} is the annual energy generation of wind energy installation. The annual wind energy generated is calculated based on the installation's capacity and assuming a capacity factor of 51\%. Recent empirical data and high-resolution modeling studies confirm that advanced offshore wind farms can achieve capacity factors of 51\% or higher \cite{An_Cai_Lu_Wang_2025a} \cite{IRENA2019}. 

\section{Results}
\subsection{Estimating the costs of installation from ORBIT model}

\begin{figure}[h!]
    \includegraphics[width=\textwidth]{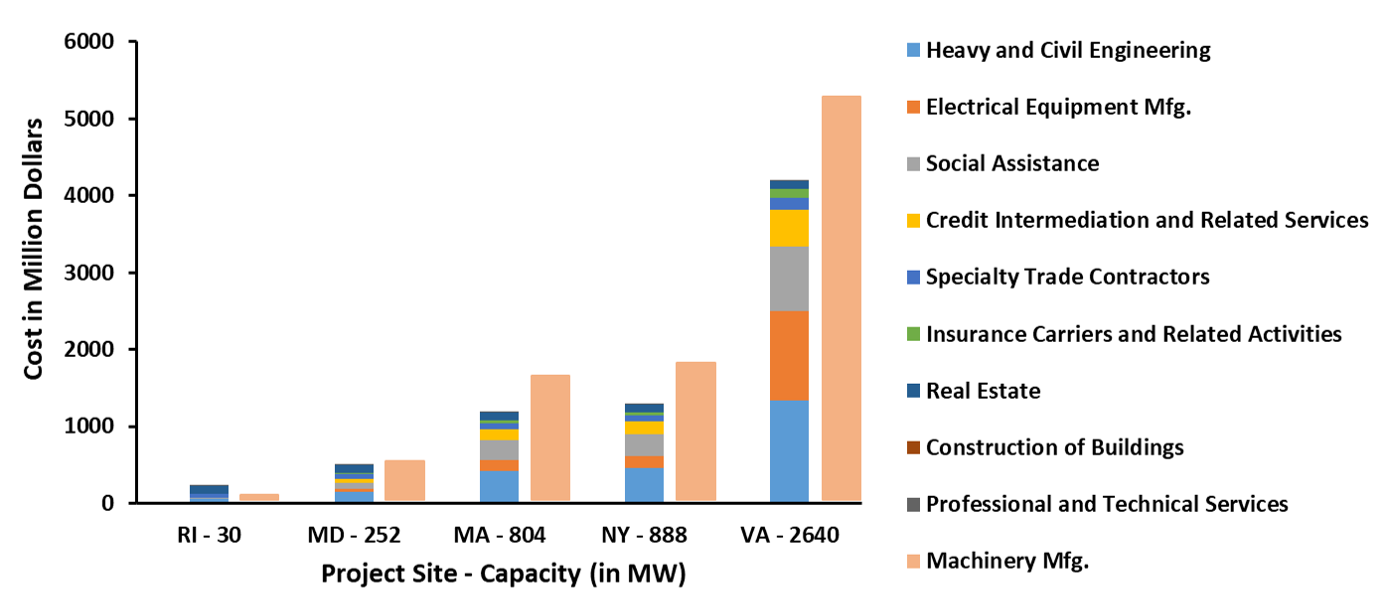}
    \caption{Project costs obtained from ORBIT}
    \label{fig:Costs}
    \centering
\end{figure}

The project construction and installation costs obtained from the ORBIT model are summarized in Figure \ref{fig:Costs} along with the sector breakdown of the total project costs for installation at each site. All cost values are expressed in nominal U.S. dollars.
The total project costs can be broadly categorized as turbine manufacturing or installation. The cost of installation per MW of installed capacity decreases with plant upsizing. the turbine manufacturing costs scale linearly with plant capacity since we assume that 12MW turbines are used for each project. The ratio of the cost of turbine manufacturing to the cost of installation increases with installed capacity. For the projects considered in this analysis, this ratio is less than one for the RI and MD projects, and greater than one for others. {Based on this trend, we could expect the cost of turbines to be the biggest project expense for large capacity projects, currently spent on imports.} In this analysis, the turbine manufacturing costs are attributed to the machinery manufacturing sector (NAICS 333). Fig. \ref{fig:Costs} summarizes the sector breakdown of the total project costs for installation at each site. Fig \ref{fig:Costs} also shows how turbine manufacturing costs outpace installation costs as project capacity increases. The details of the project cost mapping to NAICS sectors are provided in Appendix B and C.

\subsection{Multiregional economic impacts}

 \begin{figure}[h!]
    \centering
    \begin{minipage}{0.45\textwidth}
        \centering
        \includegraphics[width=1\textwidth]{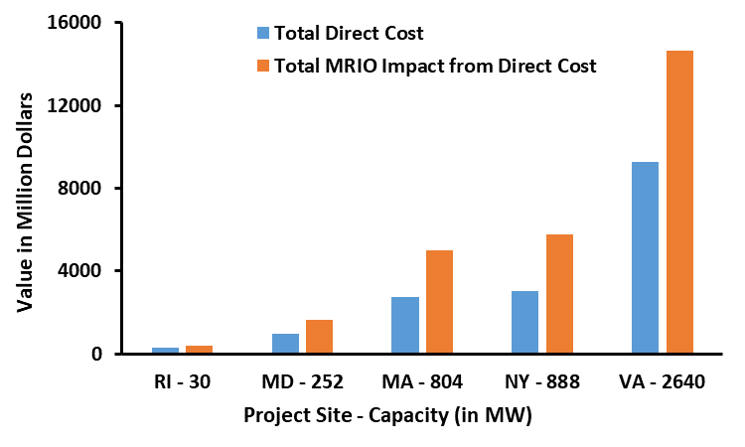}
        \caption{Total project costs and nationwide induced economic impacts}
        \label{fig:CostVsMRIO}
    \end{minipage}\hfill
    \begin{minipage}{0.45\textwidth}
        \centering
        \includegraphics[width=1\textwidth]{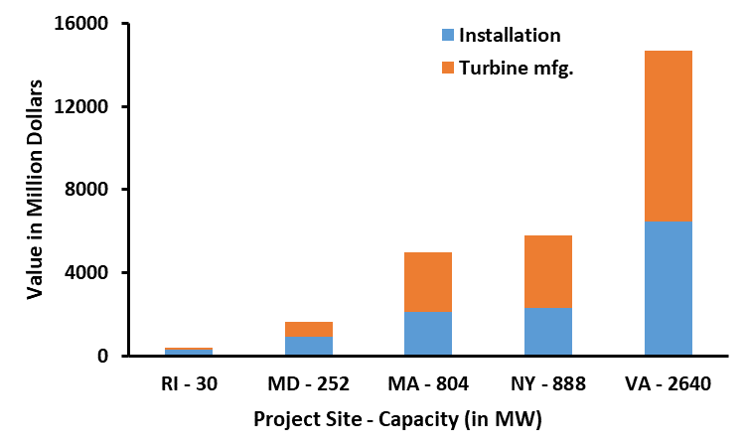}
            \caption{Nationwide MRIO impacts split by impacts from turbine mfg. and installation costs}
            \label{fig:MRIOSplit}
        \centering
    \end{minipage}
\end{figure}

The cost to install each offshore wind project for RI,  MD, MA, NY, and VA was \$296M, \$978M, \$2.8B, \$3B, and \$9.3B, respectively. The total expected economic impact induced by these projects from the MRIO analysis is \$381M,  \$1.6B, \$5B, \$5.8B, and \$14.7B, respectively. The share of total impact from non-turbine related costs is expected to be between 40-79\% for all projects, but excluding RI, the smallest project, the share of induced impact from non-turbine related costs is between 40-56\%. 
In Fig \ref{fig:CostVsMRIO}, the project cost, broken down by sector in \ref{fig:Costs}, is compared to the total economic impact induced by the project. {Fig \ref{fig:MRIOSplit} shows how much of the total economic impact induced by the project is from turbine manufacturing versus installation costs.}

\begin{figure}[h!]
    \includegraphics[width=\textwidth]{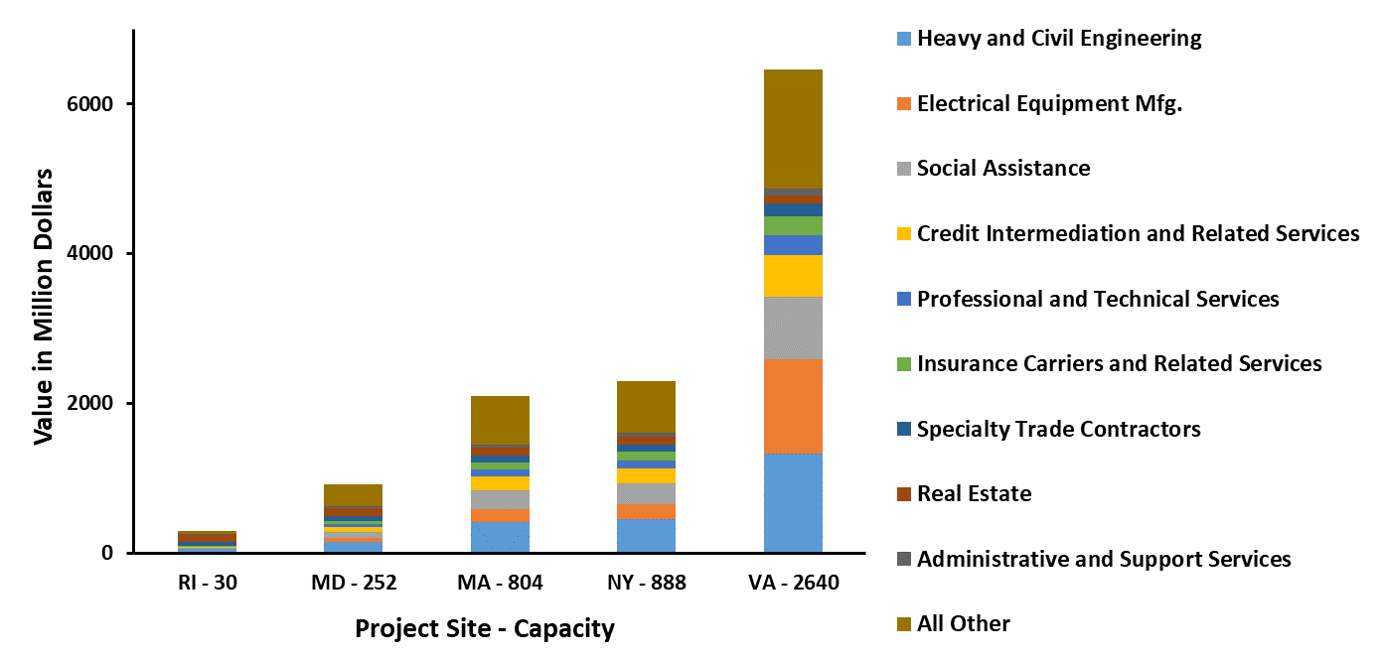}
    \caption{Nationwide Economic Impacts occurring from installation costs}
    \label{fig:InstMRIOImpact}
    \centering
\end{figure}

\begin{figure}[h!]
    \includegraphics[width=\textwidth]{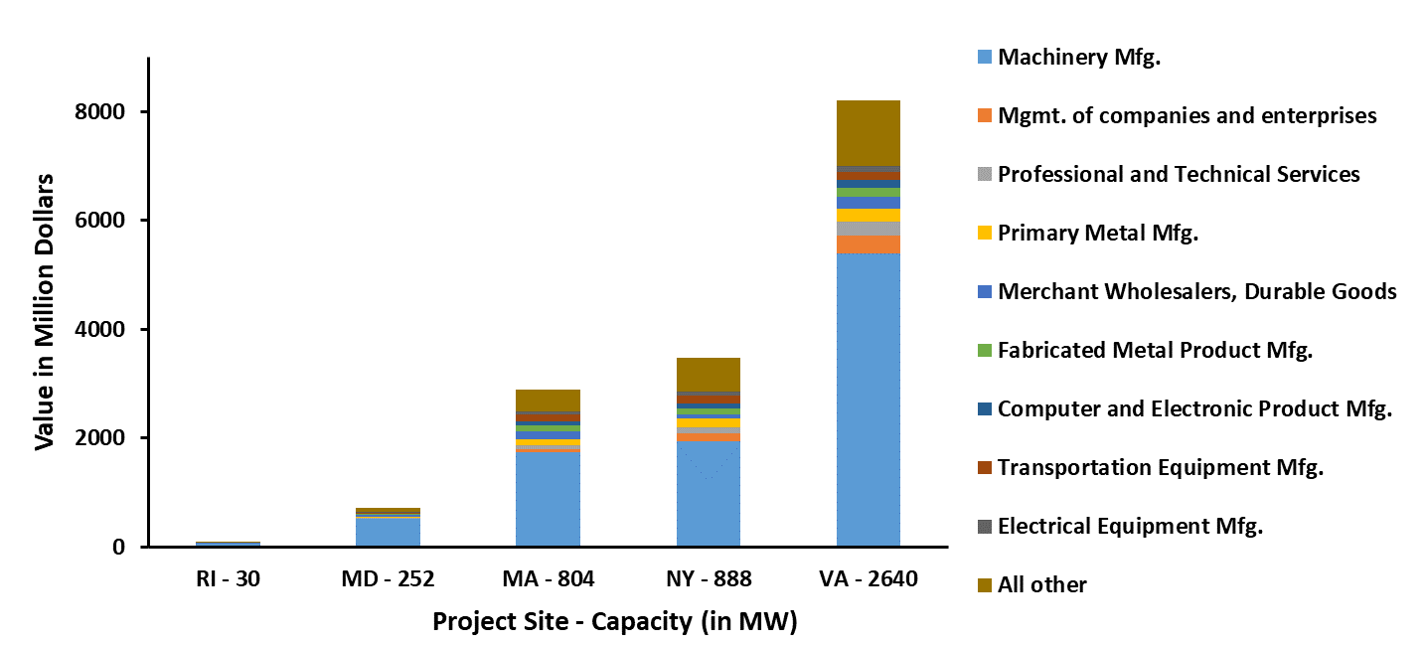}
    \caption{Nationwide Economic Impacts occurring from turbine manufacturing}
    \label{fig:TurMRIOImpact}
    \centering
\end{figure}

{As seen in Figures \ref{fig:InstMRIOImpact} and \ref{fig:TurMRIOImpact}}, the sector contributing most to the overall economic impact is NAICS sector 333 – Machinery Manufacturing, which is the sector that would manufacture the wind turbines, followed by the aggregate impact from all other sectors. The sectors contributing most to the impact of non-turbine-related project costs are NAICS sector 237 – Heavy and Civil Engineering Construction, 335 – Electrical Equipment, Appliance, and Component Manufacturing, and the combined impact from all other sectors. 

After considering the impact of Machinery Manufacturing, there is a fairly even mix of industries whose activity contributes to the total economic impact of the project, with service and manufacturing sectors represented. In contrast, materially intensive industries contribute most to emissions impacts, with most emissions due to the construction of imported turbines.

\subsection{Multiregional GHG emissions impact}
\begin{figure}[h!]
    \centering
    \includegraphics[scale=0.6]{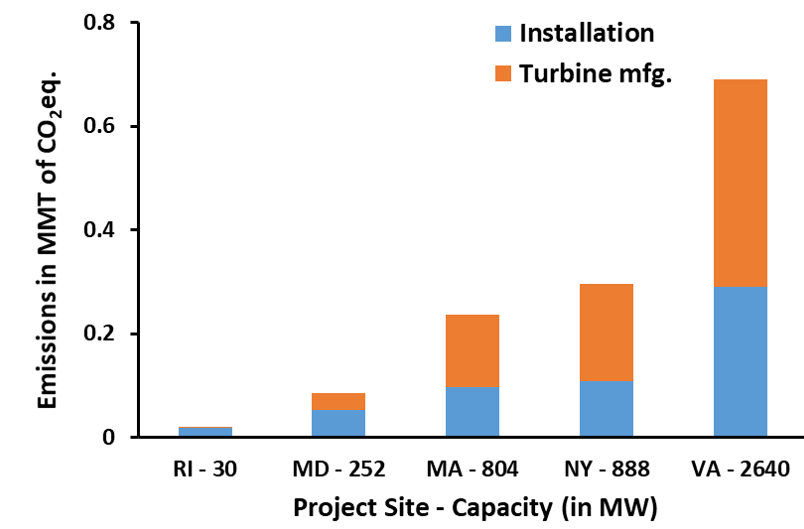}
    \caption{Total emissions split by impacts from turbine mfg. and installation costs}
    \label{fig:Emissions-Split}
    \centering
\end{figure}
The estimated nationwide emissions from each offshore wind project for RI, MD, MA, NY, and VA were 0.021 MMT, 0.085 MMT, 0.235 MMT, 0.295 MMT, and 0.689 MMT of \(CO_2\) equivalents. These emissions are a one-time occurrence that is caused during the construction and installation stage of the wind farms. The wind farms produce negligible emissions throughout their lifetime of approximately 25 years.  
Fig \ref{fig:Emissions-Split} shows the estimated emissions from turbine manufacturing and installation for all the projects considered.  As expected, the emissions occurring from turbine manufacturing and installation follow the same trend as the costs for each category.{Developing a regional emissions inventory enables us to }include the emissions in this multiregional analysis {and} to locate the sectors and {states} most responsible for emissions from domestic spending and the manufacturing sectors that will have the most significant impact on the emissions outcome of the project if they were produced domestically, even though their economic contribution is relatively small.

\begin{figure}[h!]
    \centering
    \includegraphics[width=\textwidth]{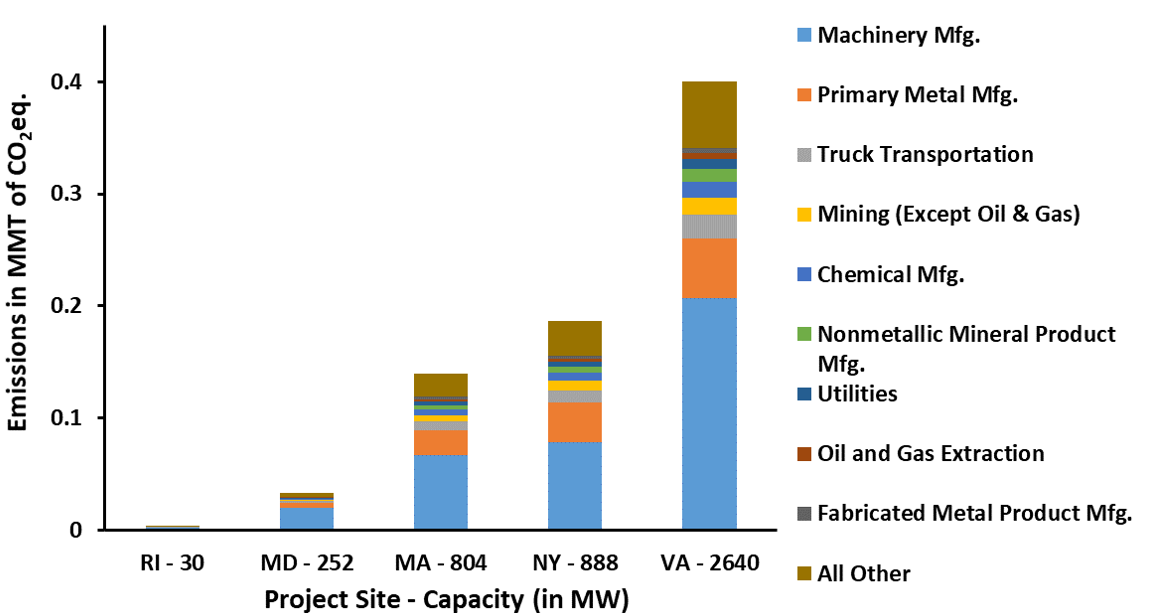}
    \caption{Total emissions occurring from turbine manufacturing}
    \label{fig:Emissions-Turb}
    \centering
\end{figure}

\begin{figure}[h!]
    \centering
    \includegraphics[width=\textwidth]{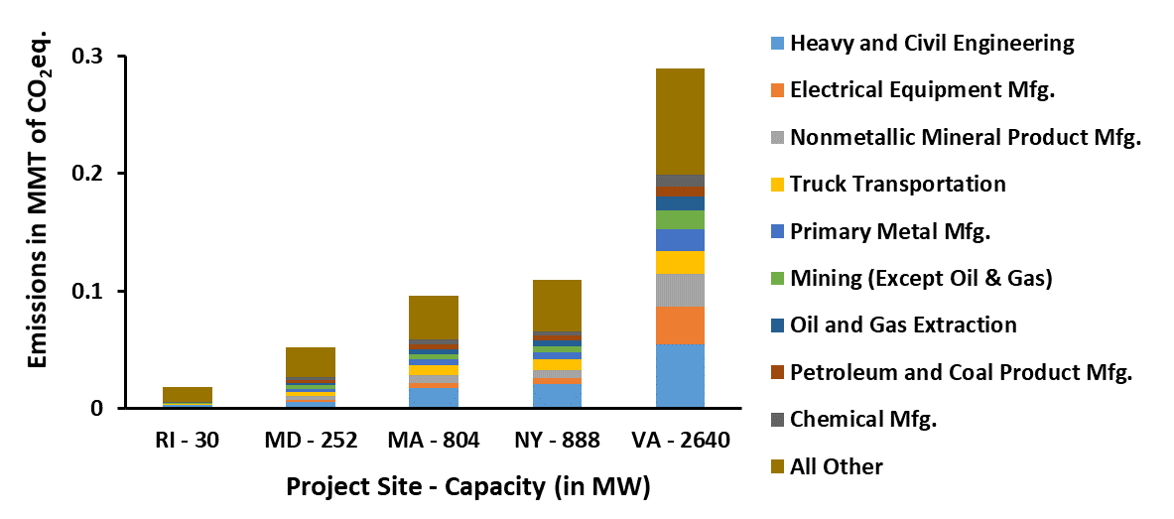}
    \caption{Total emissions occurring from installation}
    \label{fig:Emissions-Inst}
    \centering
\end{figure}

The emissions occurring in each case are split by sector to denote the top 10 industrial sectors contributing to emissions {in Figures \ref{fig:Emissions-Turb} and \ref{fig:Emissions-Inst}}. {In the case of turbine manufacturing, it is important to note that if he US started to construct turbines, while the majority of emissions would become domestic, a share from the third largest category would still be embodied in imported components because only some of the major metals used in turbines are produced in the US, but minor metals like rare earth elements used in offshore wind turbine motors would likely still be refined if not also mined in other countries.} In the case of installation, heavy engineering and construction, electric equipment manufacturing, and nonmetallic mineral product manufacturing sectors are some of the most significant contributors to emissions. The basis to develop database for regional emissions is shown in Appendix A.


\subsection{Spatial economic and emissions impact}
\begin{figure}[h!]
    \centering
    \begin{minipage}{0.48\textwidth}
        \centering
        \includegraphics[width=1\textwidth]{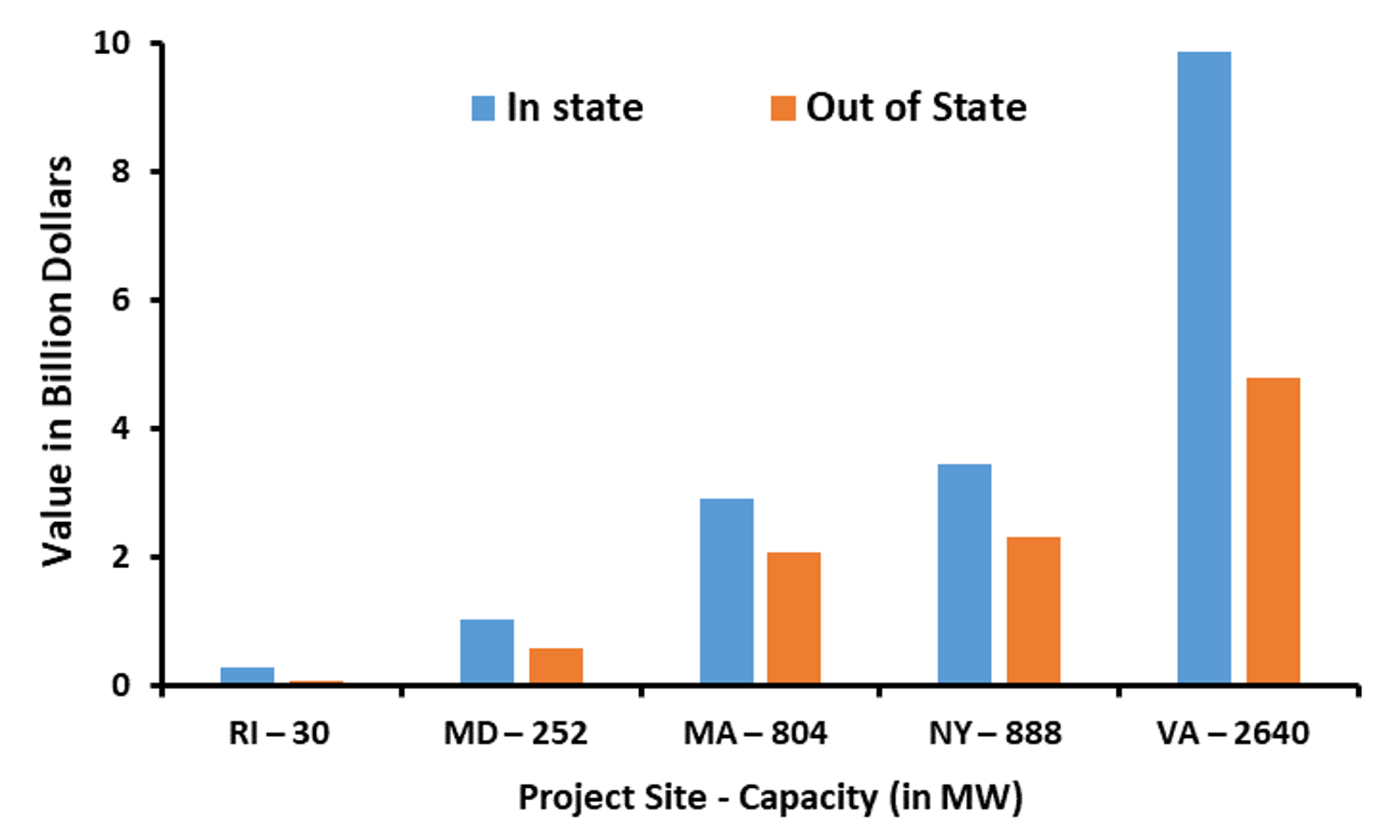}
        \caption{Economic impacts occurring within installation state and the rest of US }
        \label{fig:OutOfStateEconomicImpact}
    \end{minipage}\hfill
    \begin{minipage}{0.48\textwidth}
        \centering
        \includegraphics[width=1\textwidth]{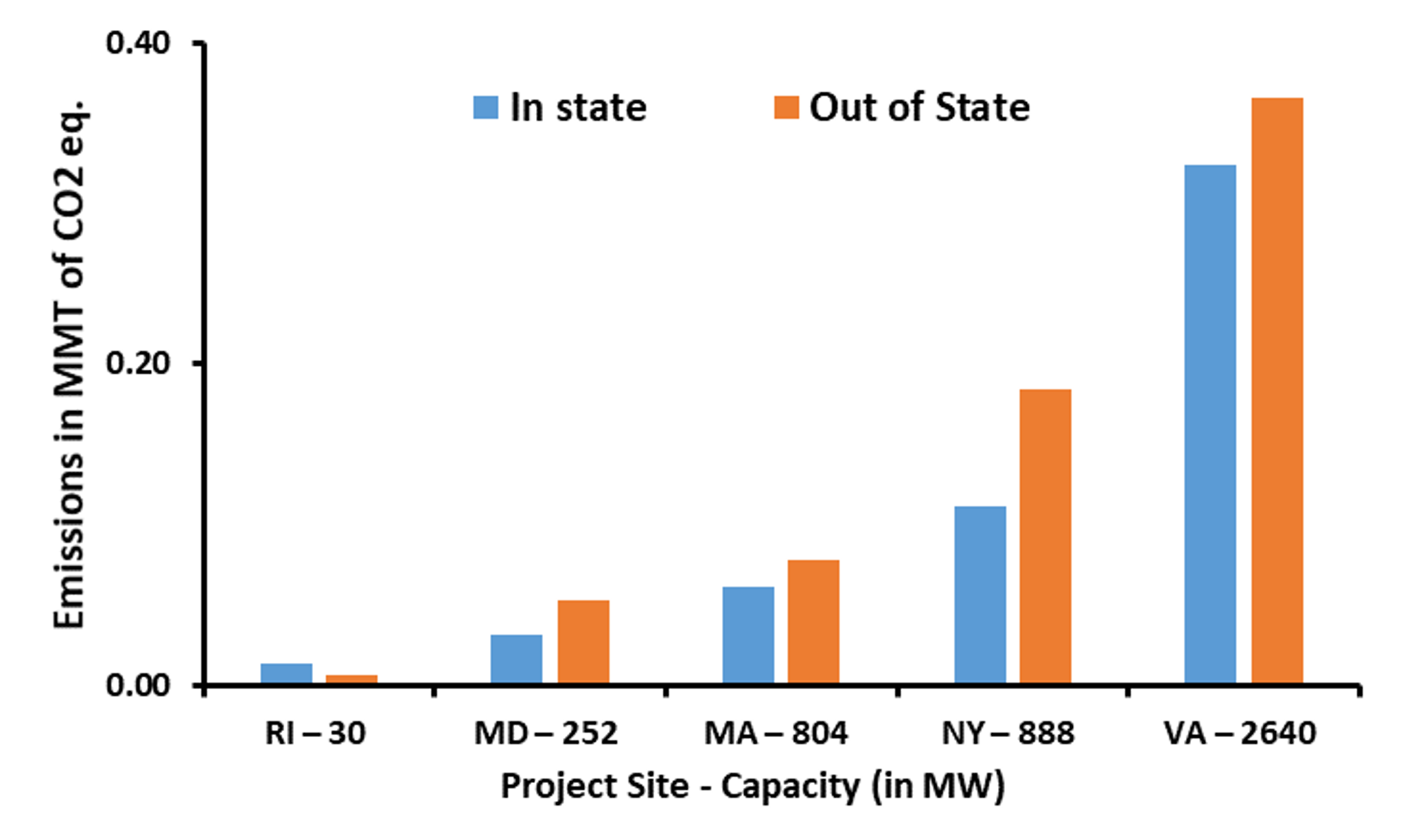}
            \caption{Emissions occurring within installation state and rest of US}
            \label{fig:OutOfStateEmissions}
        \centering
    \end{minipage}
\end{figure}

Installing offshore wind energy projects causes increased economic throughput throughout the country. There is not only a direct positive economic impact within the state of installation due to construction activities, but the spillover effects can be observed in other states that provide material and labor inputs for these projects. The induced economic activities also result in environmental impacts in these areas. Fig  \ref{fig:OutOfStateEconomicImpact} and \ref{fig:OutOfStateEmissions} show the economic and emissions impact generated within and outside of the installation state for each of the projects considered. The economic activity within the project state can be attributed to machinery manufacturing, construction, real estate, and social assistance sectors. Considering that some of these sectors, such as the social and financial assistance sectors, have lower emissions intensities, it is observed that even though there is a higher economic activity within the state compared to the out-of-state, the emissions generated within the state are lower.

\begin{figure}[h!]
    \centering
    \includegraphics[scale=0.12]{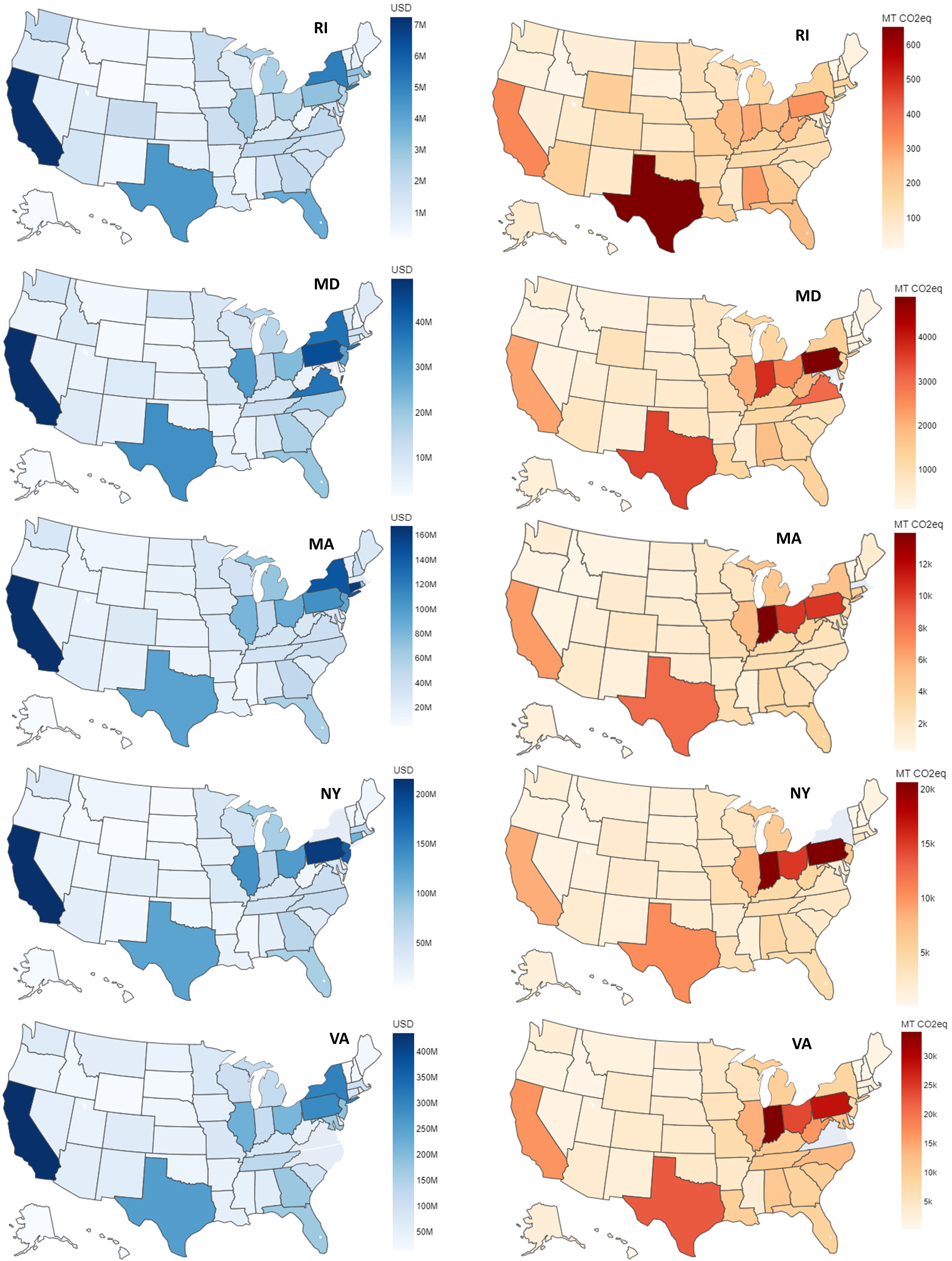}
    \caption{\textbf{Impacts occurring outside of installation state.}
    The blue maps show economic impacts and orange maps show emissions impacts.}
    \label{fig:Chloropleths}
    \centering
\end{figure}

Fig \ref{fig:Chloropleths} shows the economic impacts and emissions occurring in different regions of the US from the installation activities in project states. The maps do not include the impacts occurring within the project state. California, New York, Pennsylvania, and Texas regions experience the highest economic impacts. The economic impact occurring in California is from the Computer and Electronic Product Manufacturing and Technical, Administrative, and Management Service sectors. The impacts in Pennsylvania are observed in the primary and fabricated metal manufacturing and machinery manufacturing sectors. Impacts in Texas are due to the Professional, Scientific, and Technical Services, and Fabricated Metal product manufacturing sectors, and impacts in New York are due to Finance and Insurance sectors. The sectors that see a high economic impact in states of California and New York are generally less material and emissions intensive. Hence, it is observed that although these states experience a high economic impact, there is a much lower emissions impact occurring in these states.

The primary metal manufacturing, non-metallic mineral product manufacturing, oil and gas extraction and utilities sectors are the largest emitters for each project. This trend is expected since these are the sectors that provide a majority of the material input to the turbine manufacturing and installation operations. The turbine manufacturing costs are mapped to the Machinery Manufacturing sector, and the turbines are predominantly made from steel. Indiana, Ohio, Pennsylvania, and Texas are the four regions that are most impacted in terms of emissions. Indiana, Ohio, and Pennsylvania regions have some of the nation's largest steel industries, which are the major cause of emissions in these regions. Emissions occurring in Texas can be attributed mainly to the state's Oil and Gas Extraction industry and Truck Transportation. {Decarbonizing primary metal manufacturing and related heavy industries} in these states such as Indiana, Ohio, Pennsylvania and Texas will {be necessary} to lower emissions from building off-shore wind projects along with stimulating the economy of these regions. Increased demand for manufacturing, construction and finance in these regions will lead to more employment opportunities that we have not quantified.

In the case of the Rhode Island and Maryland projects, the installation cost is  higher than the cost of turbine manufacturing. This is because the capacity of these projects is smaller and they use fewer turbines. These projects are therefore also less steel intensive than the other three projects and this trend can be observed in the spatial emissions map. A more significant portion of the emissions caused by these projects can be attributed to the Oil and Gas Extraction, Utilities, and Nonmetallic Mineral Product Manufacturing sectors. Hence, more of the emissions from these two smaller projects come from Texas and Pennsylvania instead of the states that would provide steel for turbine manufacturing. The economic and emissions impacts for each project are tabulated in the Supplementary Information.

\subsection{Payback period analysis}
 
The economic payback period for each project is shown in Table \ref{tab:EconPayback}. This was calculated using the method described in section \ref{CBP}. The cost of the projects (C\textsubscript{i}) is the total project cost estimated by the ORBIT model. AEP is the annual power generation calculated using the project capacity and a capacity factor of 0.51. P\textsubscript{s} is the average cost of electricity for each state in which the project is installed \cite{EIAElectricityPrice}. C\textsubscript{op} is the annual operations cost obtained from the ORBIT model. The economic payback period for these projects is between 5-15 years. These estimates are also comparable with those for specific solar installations \cite{Chandel2014Techno-economicCity}. Generally, projects with a shorter payback period are preferred over those with more extended payback periods. There is no direct correlation between the payback periods and the plant capacity. However, the projects with either very low or high capacities are less economically favorable from a payback period perspective. However, all commercially sized projects considered
would recover their original investment cost in at most about half of their lifetime while creating other positive economic spillover effects.

\begin{table}[]
\resizebox{\columnwidth}{!}{%
\begin{tabular}{|c|c|}
\hline
\textbf{Project Site - Capacity (MW)} & \textbf{Payback period (Years)} \\ \hline
RI - 30                               & 15.2                            \\ \hline
MD - 252                              & 11.4                            \\ \hline
MA - 804                              & 5.1                             \\ \hline
NY - 888                              & 6.6                             \\ \hline
VA - 2640                             & 13.6                            \\ \hline
\end{tabular}%
}
\caption{Economic payback period for the projects of interest}
\label{tab:EconPayback}
\end{table}

Table \ref{tab:EmissionsPayback} shows the carbon payback period for each offshore wind energy project considered in this study. The carbon payback period calculation is explained at the end of the methods section. The payback period considers the projected trend of decarbonization of the energy sector in the US \cite{USEIAProjection}. The emissions for energy generation from each source are calculated from the emissions factors reported in the literature \cite{offshorewindadvisory}. R\textsubscript{avgelec} is the emissions intensity of conventional energy generation considering
future decarbonization. These five projects will offset the emissions caused during turbine manufacturing and installation within a year of operation, which is significantly less than the projects' lifespan. The carbon payback periods for these projects are comparable to other existing studies; most of which use process-based life cycle assessment (LCA) to estimate the emissions generated. For example, an LCA study of a German wind farm reports a carbon payback period of about nine months \cite{Wagner2011LifeVentus}. Most of the studies presented in the literature report emissions payback values to be less than one year \cite{Fonseca2022GreenhouseNortheast,Rankine2006EnergyTurbine}. The minor variation in these values can be attributed to different life cycle boundaries and datasets used.

\begin{table}[]
\resizebox{\columnwidth}{!}{%
\begin{tabular}{|c|c|}
\hline
\textbf{Project Site - Capacity (MW)} & \textbf{Payback period (Months)} \\ \hline
RI - 30                               & 6                                \\ \hline
MD - 252                              & 3                                \\ \hline
MA - 804                              & 2                                \\ \hline
NY - 888                              & 3                                \\ \hline
VA - 2640                             & 2                                \\ \hline
\end{tabular}%
}
\caption{Carbon payback period for the projects of interest}
\label{tab:EmissionsPayback}
\end{table}

\section{{Discussion}}
Although wind energy is one of the cleaner energy generation sources, turbine manufacturing and project installation produce significant emissions. However, this is a one-time consequence occurring during installation. The project will be functional for about 25 years post-installation, and the emissions occurring during this time are negligible compared to the installation emissions. Based on this analysis, all projects considered will offset emissions caused during the turbine manufacturing and installation phase within a year of operation, significantly less than the projects' lifetimes. This highlights the significance of these renewable energy projects in meeting climate mitigation targets. The energy and investment spent in setting up wind plants are usually redeemed in a relatively short period compared to their lifetime. Additionally, offshore wind plants 
reduce dependence on fossil fuels and reduce the vulnerability of nations to fossil fuel supply disruptions and price shocks. On the other hand, particularly for offshore turbines that use rare earth permanent magnets, the decreased dependence on fossil fuels comes with an increased dependence on the availability of critical metals.


There are several important limitations to this study we would like to acknowledge. While our impact estimates include all upstream economic and emissions supply chain impacts we have not included impacts from end-of-life treatment. 
The waste generated at the end-of-life of these projects is of concern, but several LCAs have found this life cycle stage to be the least impactful \cite{alsaleh_comprehensive_2019}; in the case of an analysis of onshore turbines in Texas, the benefits from recycling turbine components even lead to the impact to global warming from the end-of-life phase being negative \cite{alsaleh_comprehensive_2019}. Standard approaches for recycling offshore wind turbines are still in development and we chose to exclude this phase, which in other studies contributed little to overall impact, to avoid introducing major uncertainty into our impact estimates. 
Over 90\% of the turbine {can be} made from recyclable materials like steel, copper, and recyclable plastics. The turbine blades are typically made from polymers and fiberglass, which are not entirely recyclable. However, a recycling process that converts these polymers to syngas and recovers high-purity fiberglass is now commercially  available, which diverts a significant amount of waste that would otherwise end up in the landfill to a productive use \cite{WindTurbineRecycling}. More solutions to upcycle end-of-life turbines and proactive planning for their end of life are needed, but should happen concurrently with deployment. 
Potential future work in this regard includes integration of dynamic modeling frameworks, such as the Circular Economy Lifecycle Assessment and Visualization (CELAVI) tool. This tool captures the evolution of environmental impacts as supply chains become more circular over time. CELAVI is suitable to evaluate non-steady-state systems, where recycling pathways and infrastructure are still under development \cite{Hanes_Ghosh_Key_Eberle_2021}. Additionally, using scenario-based LCA approaches can help model the environmental trade-offs of alternative end-of-life options such as mechanical recycling, solvolysis, and component reuse to evaluate how new end-of-life technologies and circular design strategies may influence the long-term sustainability of offshore wind systems \cite{Alavi_Khalilpour_Florin_Hadigheh_Hoadley_2025}. EOL planning must be incorporated in the early stages of design to minimize the overall impacts of offshore wind projects.

Another contribution to project lifecycle impacts that has been excluded but would negatively increase costs and add to emissions is the externality cost or social cost of carbon that could be applied to each project \cite{us_epa_social_nodate}. Including externality impacts would make the estimation of indirect economic costs and thereby project impacts even more complete. For the social costs of carbon, the cumulative social cost would be added to the numerator in Eq. \ref{eq2}, thereby increasing the payback period. While a rigorous accounting of these costs were outside the scope of this analysis, additional discussion of how these costs could be incorporated in future analysis in available in Appendix D. Another limitation is that we assume turbines are manufactured domestically in order to estimate the impacts from an activity that is intended to happen domestically in the future. The economic linkages between international sectors that currently manufacture turbines are likely somewhat different as are the sectoral emissions impacts between different countries. Using a multiregional MRIO model of different world regions would allow for a more accurate accounting of the regional emissions incurred in producing offshore wind turbines for the next several years. However, we were interested in capturing the impacts from future domestic manufacturing that would ideally be available for the majority of offshore wind projects going forward. Finally, this analysis does not account for the cost and emissions of energy transmission and storage systems that will also be needed to integrate offshore wind energy plants with the electricity grid. Quantifying and allocating a portion of the grid infrastructure build-out required to accommodate renewables to offshore wind energy was outside the scope of this work. However, the required infrastructure and associated impacts for grid upgrades are less visible but still important contributors to overall material requirements and associated impacts of the transition to cleaner energy. 




One of the sustainable development goals adopted by all United Nations member states is to ensure access to affordable, reliable, sustainable, and modern energy for all by 2030. There has been continued advancement towards this goal, but the progress is not happening fast enough \cite{The-Sustainable-Development-Goals-Report-2023.pdf}. To accomplish this goal, there is a need for increased investments in renewable energy, accelerating the rate of electrification, and adopting more effective and comprehensive policies to promote renewable energy \cite{IEA201320Recommendations}. Adopting renewable technologies also creates new jobs in various sectors, contributing to local economies and providing a gainful way to transition to a low-carbon economy. A closely related goal is to take action to combat climate change and its impacts. Rapid and sustainable reduction in emissions through this decade is critical to limit global warming 1.5$^{\circ}$C above pre-industrial levels. Fossil fuels are the largest contributors to climate change globally, and offshore wind is a viable alternative to reduce dependence on fossil fuels and 
invest in energy that is cleaner and affordable \cite{UnitedNations}. Switching to energy generation with no use phase emissions also helps address air pollution and reduce threats from it to general health and well-being.  

Aside from GHG emissions, there are other potential environmental and social impacts due to the adoption of newer renewable energy sources worth considering. Pertinent potential environmental impacts include changes in land use and disruption to local wildlife and ecosystems. Offshore wind installations do not have much effect on land use. However, the degree to which they may impact marine ecosystems, fisheries, and ocean and atmospheric dynamics is only starting to be understood \cite{akhtar_impacts_2022}. The noise and vibrations produced during the construction phase, particularly during the installation of the foundation, are said to cause physical injuries in several species of marine animals. The noise during the operation phase is not loud enough to cause physical harm to marine animals. \cite{WingGoodale2019AssessingStates}. Other studies have investigated the effect of wind farm siting on various seabird guilds \cite{Mooney2020AcousticFishery}. There are only a limited number of such studies presented in literature since this is an emerging technology, and its long-term impacts have not been evaluated thus far.
 One study specific to the Northeast US evaluated the expected impacts of offshore wind on the marine ecosystems and fishery activities off the Atlantic coast \cite{methratta_offshore_2020}. They conclude that the stressors of wind energy may help and harm in different ways but more work is urgently needed to understand the cumulative impact. Also, they acknowledge that there are great challenges to enabling offshore wind to coexist with fishing activities, but that the challenges are not insurmountable with cross-sectoral and regional collaboration and innovation in research design related to surveying and fisheries management \cite{methratta_offshore_2020}.
Further, there is a growing concern that offshore wind plants can potentially reduce the attractiveness of the landscape and negatively impact the tourism industry. However, recent studies show that offshore wind plants, especially those located  20 miles offshore, have little negative effect on the tourism industry \cite{Parsons2020TheData}. Any likely reduction in employment in the tourism industry will be outweighed by increased employment in the wind power industry. There is also evidence of a potentially considerable number of curiosity trips to view these newly installed projects \cite{Trandafir2020HowFarm}. There are many factors that should go into evaluating the sustainability of any endeavor and for offshore wind, factors like marine ecological and local livelihood disruptions and supply chain pressure from increased demand for advanced energy materials to build turbines are examples of important factors that should influence where, how many wind farms a place should endeavor to build, and how fast to build them. A more comprehensive sensitivity and uncertainty analysis could further strengthen the results. However, conducting such an analysis requires access to detailed empirical data, which is currently unavailable. Without this data, the findings have limited applicability to real-world wind energy deployment. We recommend addressing these aspects in future work, once the necessary data is available to enable a more targeted analysis. Evaluating the potential synergies and tradeoffs presented by these factors is a challenge because the impacts are a) of different qualities and b) are just starting to be understood. Provided the attributes corresponding to each impact can be quantified to a similar degree that GHG emissions can currently be quantified, frameworks like multi-criteria decision analysis (MCDA) could be used for incorporating these diverse factors into decision-making about offshore wind projects, as has been done to weigh technical considerations of offshore wind projects \cite{wang_offshore_2022} and to decide on preferred ecological conservation strategies \cite{thompson_stakeholder_2019}. To operationalize such a tool, the impacts themselves need to be quantified as well as the weighting factors used to normalize them, often requiring consensus or at least survey-based data \cite{diaz_novel_2022}. These are among the factors influencing sustainability that deserve more consideration in research and policymaking and possible ways to address them.


\section{{Conclusions}}
This study presents a quantitative analysis of the regional economic and environmental impacts of upcoming offshore wind projects within the US. This is done using environmentally extended multiregional input-output analysis. {The major methodological contribution of this work is the development of a state-level EEIO GHG emissions dataset compatible with the US IELab and other subnational IO models of the US. Emisions accounting at state level is necessary to identify where emissions hotspots are within the country that would be triggered by offshore wind energy development.} The results from this analysis demonstrate that introducing new offshore wind plants will generate a sizeable positive economic impact across the country and will offset the emissions attributed to their creation within about a year of operation. 


Perhaps the most accessible indicator of sustainability for a project directly aimed at curbing GHG emissions, but again, not the only one that matters, is one that shows to what degree the project will actually reduce emissions given that its creation will induce significant emissions in the process. This work seeks to apply such an indicator, {carbon payback period,} to planned US offshore wind plants and finds that from an economy-wide perspective, the climate change mitigation potential of these projects do far outweigh the upfront emissions burden. Addressing other important stressors to Earth and social systems that could be created by expanding offshore wind was outside the scope of this work, but will hopefully be considered along with offshore wind's GHG emissions mitigation potential as the US begins its transition to an economy more in line with Earth systems.

\vspace{0.5cm}

\textbf{Declaration of competing interest}

The authors have no conflicts of interest to declare.

\vspace{0.25cm}

\textbf{Acknowledgements:}

Authors are grateful for support from the U.S. National Science Foundation CBET-1805741 and FMRG- 2229250. We also thank Manfred Lenzen and Arne Geschke for their support to develop US IE Lab models. Authors would like to thank the reviewers for their feedback that has helped improve this manuscript.

\vspace{0.25cm}

\textbf{Author contributions:}

\textbf{Apoorva Bademi:} Conceptualization, Methodology, data curation, Formal analysis, Visualization, Writing - original draft; \textbf{Miriam Stevens:} Conceptualization, Methodology, Software, Formal analysis, Data curation, Writing - Original Draft, Visualization; \textbf{Isha Sura:} Conceptualization, Methodology, Data curation, Writing - Original Draft; \textbf{Shweta Singh:} Conceptualization, Methodology, Writing - review \& editing, Supervision, Project administration, and Funding acquisition

\vspace{1cm}


\newpage
\appendix

\section{Regional GHG database}
\label{sec:GHGMethod:appendix}
\begin{figure}[h!]
    \includegraphics[width=0.9\textwidth]{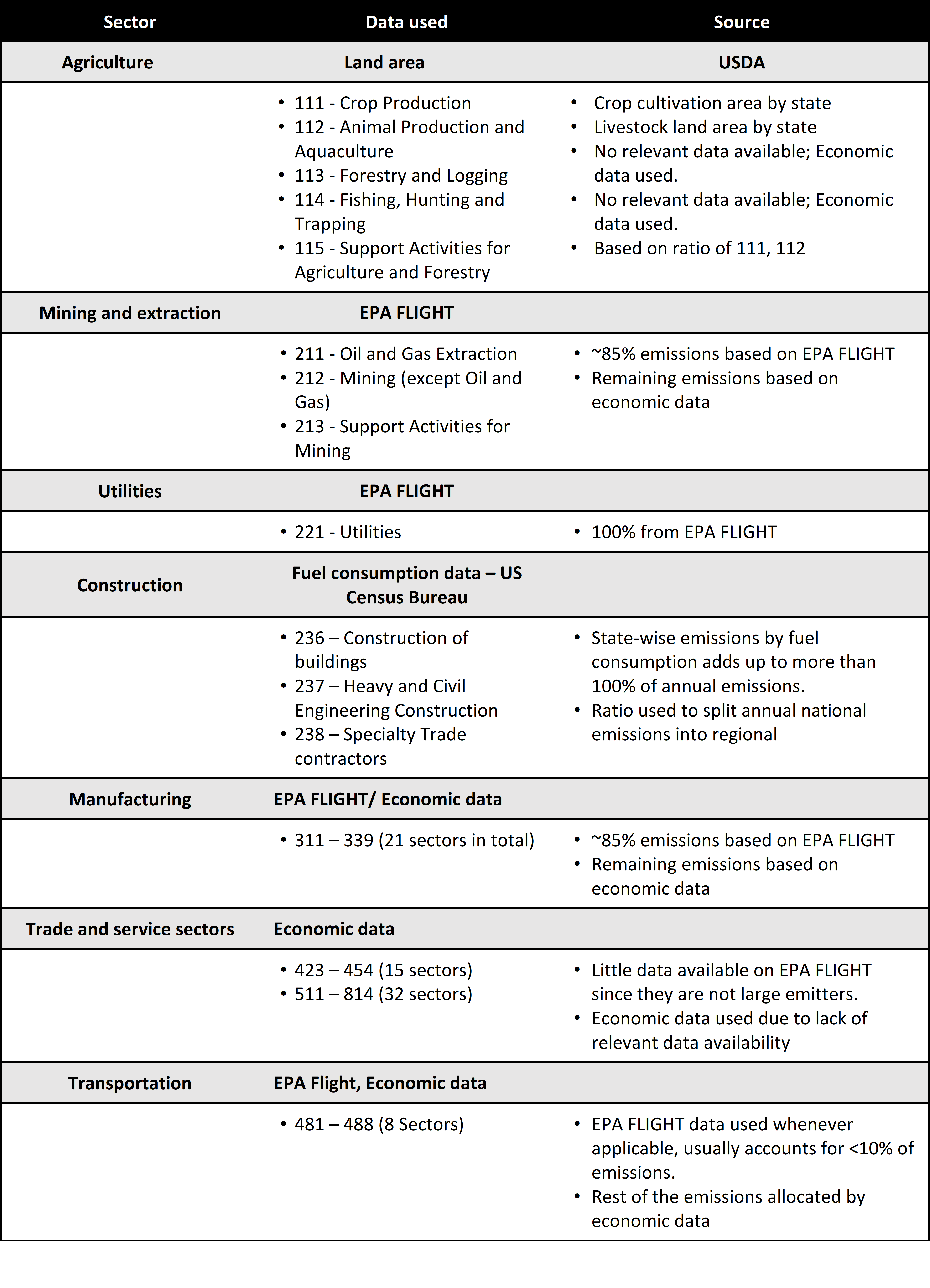}
    \caption{Methodology for regionalizing emissions for different sectors}
    \label{fig:GHGDisagg}
    \centering
\end{figure}
The national level GHG inventory, published by the US EPA, is mapped to 397 industrial sectors as defined by the US BEA, along with 20 personal and federal sectors. This database exists for the years between 2011 and 2016 \cite{Azuero-Pedraza2022IncorporatingModels}. The national emissions for the year 2017 were obtained using the same method. The Inventory of U.S Greenhouse Gas Emissions and Sinks: 1990-2020, published in 2022, was used as the primary data source. The BEA sectors were then mapped to corresponding NAICS sectors since this data is eventually combined with economic data at a NAICS level. The national data is then disaggregated into regional data using the most relevant spatial data available for each industrial sector. The final dataset reports emissions for 100 industrial sectors at a 52-region level, including the 50 US states, DC, and Puerto Rico. Since each economic group is distinct in operation, no single dataset can be used to disaggregate the data into resolved sectors. Hence, the most relevant data for each major category is used. This split is based on economic data when no pertinent information is available. The primary data for disaggregation was obtained from the US EPA Facility Level Information on GHG Tool (EPA FLIGHT) \cite{EPAFLIGHT}. The EPA FLIGHT provides information about greenhouse gas emissions from large facilities across the US. The EPA defines large emitters as any facility producing over 25,000
metric tons of CO2 equivalents yearly. This reports data regarding location (City, State, Zip code),
NAICS code and Industry type for each facility. This is an excellent tool for regionalization since
facilities can be easily mapped to a state, and these emissions can be added up across states to
obtain regionalized emissions. However, since this data set only considers large emitters, it only
accounts for 85 to 90\% of the total annual emissions across all industries. The 10-15\% emissions must be accounted for and allocated to regions. This is also done using the most relevant information available for each sector, but economic data is used in cases where no pertinent data is available. This dataset can be improved using more appropriate data for disaggregating service sector emissions. \ref{fig:GHGDisagg} below summarizes the different data sources used for regionalization.

\newpage
\section{ORBIT Cost mapping}
\label{sec:ORBITCostMapping:appendix}

\begin{landscape}
\begin{table}[]
\begin{tabular}{|l|l|l|l}
\cline{1-3}
\textbf{Cost category}           & \textbf{NAICS description}                                              & \textbf{NAICS} &  \\ \cline{1-3}
Array System                     & Communication and energy wire and cable manufacturing                   & 33592          &  \\ \cline{1-3}
Export System                    & Communication and energy wire and cable manufacturing                   & 33592          &  \\ \cline{1-3}
Offshore Substation              & Substation and switching station, power transmission line, construction & 237130         &  \\ \cline{1-3}
Scour Protection                 & Other heavy and civil engineering construction                          & 237990         &  \\ \cline{1-3}
Substructure                     & Wind Turbine Manufaturing                                               & 333611         &  \\ \cline{1-3}
Array System Installation        & Cable Laying                                                            & 237130         &  \\ \cline{1-3}
Export System Installation       & Cable Laying                                                            & 237130         &  \\ \cline{1-3}
Offshore Substation Installation & Substation and switching station, power transmission line, construction & 237130         &  \\ \cline{1-3}
Scour Protection Installation    & Other heavy and civil engineering construction                          & 237990         &  \\ \cline{1-3}
Substructure Installation        & Structural Steel and Precast Concrete Contractors                       & 238120         &  \\ \cline{1-3}
Turbine Installation             & Heavy and civil engineering construction                                & 237            &  \\ \cline{1-3}
Turbines                          & Wind Turbine Manufaturing                                               & 333            &  \\ \cline{1-3}
Insurance                  & Insurance agencies and brokerages                                       & 524210         &  \\ \cline{1-3}
Financing                  & Credit Intermediation and related activities                            & 522            &  \\ \cline{1-3}
Contingency                & Emergency and other relief services                                     & 624230         &  \\ \cline{1-3}
Commissioning              & Building commissioning services                                         & 541350         &  \\ \cline{1-3}
Decommissioning            & Site preparation contractors                                            & 238910         &  \\ \cline{1-3}
Site Auction           & Real Estate                                                             & 531            &  \\ \cline{1-3}
Site Assessment        & Site preparation contractors                                            & 238910         &  \\ \cline{1-3}
Construction Plan      & Construction of buildings                                               & 236            &  \\ \cline{1-3}
Installation Plan      & Professional, scientific and technical services                         & 541            &  \\ \cline{1-3}
\end{tabular}
\caption{Mapping costs from ORBIT cost categories to NAICS Codes}
\label{tab:CostMapping}
\end{table}
\end{landscape}

\newpage
\section{ORBIT Cost category description}
\label{sec:ORBITCostCategoryDescription:appendix}

\begin{landscape}
\begin{table}[]
\begin{tabular}{|l|l|}
\hline
\multicolumn{1}{|c|}{\textbf{ORBIT Cost category}} &
  \multicolumn{1}{c|}{\textbf{Cost category description}} \\ \hline
Array System                     & Turbine array cabling system design                                        \\ \hline
Export System &
  Export cabling system (from offshore substation to land-based grid connection) design \\ \hline
Offshore Substation              & Connects array cable system to export cable system                         \\ \hline
Scour Protection                 & Necessary scour protection material for a fixed turbine base               \\ \hline
Substructure                     & Construction of the wind turbine base struture sitting below the waterline \\ \hline
Array System Installation        & Installation costs for the turbine-to-turbine cable system                 \\ \hline
Export System Installation       & Installation costs for the export cable system                             \\ \hline
Offshore Substation Installation & Installation costs of the offshore substation                              \\ \hline
Scour Protection Installation    & Installation costs of the scour protection material                        \\ \hline
Substructure Installation        & Installation costs of the turbine substructure                             \\ \hline
Turbine Installation             & Installation costs of the offshore turbine                                 \\ \hline
Turbine                          & Manufacture of the offshore turbine topside components                     \\ \hline
Soft Costs                       & Non-mechanistic soft costs dependent on project size                       \\ \hline
soft\_insurance                  & Default, scaled with project size                                          \\ \hline
soft\_financing                  & Default, scaled with project size                                          \\ \hline
soft\_contingency                & Default, scaled with project size                                          \\ \hline
soft\_commissioning              & Default, scaled with project size                                          \\ \hline
soft\_decommissioning            & Default, scaled with project size                                          \\ \hline
Project Development &
  \begin{tabular}[c]{@{}l@{}}Upfront permitting, review, and engineering steps required before construction\end{tabular} \\ \hline
project\_site\_auction           & Default model value                                                        \\ \hline
project\_site\_assessment        & Default model value                                                        \\ \hline
project\_construction\_plan      & Default model value                                                        \\ \hline
project\_installation\_plan      & Default model value                                                        \\ \hline
\end{tabular}
\caption{ORBIT cost category descriptions. For details on calculation of each categorical cost, refer to the official documentation on the ORBIT model \cite{nunemaker_orbit_2020,nunemaker_github_nodate}.}
\label{tab:CostCategoryDescription}
\end{table}
\end{landscape}

\newpage

\newpage
\section{Overview of social cost of carbon for inclusion in future analyses}
\label{sec:SocialCostCarbon:appendix}

\begin{center}
Considerations for Including the Social Cost of Carbon (SC-CO2)\\
in the Economic Payback Period (PB) \\
for Offshore Wind Energy Impacts
\end{center}

The US has developed a set of social cost of carbon (SC-CO2) values for use in government cost benefit analysis. While outside the scope of our main analysis, these values could be used when estimating payback periods to include externality costs from the project installation and operation emissions. This would make the estimation of indirect economic costs and thereby project impacts even more robust for future analyses. Inclusion of such costs would make for a tiered approach for estimating economic and GHG emissions impacts of cleaner energy projects.

We have referenced the EPA method for calculating SC-CO2 from 2016 \cite{us_epa_social_nodate}. For each emission year, 4 values of SC-CO2 are recommended. 3 values are based on average SC-CO2 from 3 integrated assessment models (IAMs) run with discount rates at 2.5\%, 3\%, and 5\%. The 4th value represents the marginal damages associated with the lower probability, but high-impact outcomes of climate change (extremely bad but unlikely effects). It is the 95th percentile value for SC-CO2 based on a 3\% discount rate.

A more moderate estimation and one including more climate risk would include two calculations including SC-CO2, one using the 3\% discount rate and the other using the 95th percentile value with 3\% discount rate.

Source for SC-CO2 values: 
Table A1: Annual SC-CO2 Values: (2007\$/metric ton CO2) in Technical Support Document: Technical Update of the Social Cost of Carbon for Regulatory Impact Analysis Under Executive Order 12866 \cite{us_epa_social_nodate}.

The current economic payback period ($EPB$) is calculated:

\begin{equation}
EPB = \frac{C_i}{(AEP*P)-C_{op}} 
\end{equation}
\\
C\textsubscript{i} is the cost of initial investment, AEP is the net annual energy generation, P is the average cost of electricity, and C\textsubscript{op} is the cost of yearly operation as obtained from the ORBIT model. The denominator represents the net revenue generation from the plant.

Including SC-CO2, the $EPB$ calculation would be:

\begin{equation}
EPB = \frac{C_i + \text{SC-CO2}_{project}}{(AEP*P)-C_{op}} 
\end{equation}
\\
where $\text{SC-CO2}_{project}$ is the total social externality costs of emissions generated by the project.\\
\\
\begin{flushleft}
$\text{SC-CO2}_{project}$
\end{flushleft}

\begin{flushleft}
\hspace{10pt}= \text{SC-CO2} from project installation emissions\\ 
\hspace{20pt} + \text{SC-CO2 from operation emissions}
\end{flushleft}

\begin{flushleft}
\hspace{10pt} = $SC-CO2_{project-installation-year}$*$\Delta e$ \\  
\hspace{20pt} +  $\sum_{y=1}^n\text{SC-CO2}_{\text{project-installation-year}}$*\text{annual operational emissions in n}
\end{flushleft}

\begin{flushleft}
where $\Delta e$ is the value of direct and indirect emissions from installation from the EEIO analysis
\end{flushleft}



\newpage

 \bibliographystyle{elsarticle-num} 
 \bibliography{cas-refs}





\end{document}